\documentclass[preprint]{elsarticle}
\usepackage{bm}
\usepackage[T2A]{fontenc}
\usepackage[cp1251]{inputenc}
\usepackage{amsfonts,amssymb,amsmath}
\usepackage{amsmath,graphicx}
\usepackage{graphicx,pifont} 
\usepackage{esdiff}
\usepackage{epsfig}
\usepackage{color}

\begin{document}
	\title{Field representation of interatomic interactions: from relativistic dynamics to microscopic thermodynamics of both many-body and few-body systems}
	
	\author{A.~Yu.~Zakharov~$ ^{1} $,  V.~V.~Zubkov~$ ^{2} $}
	
	\address{$ ^{1} $Yaroslav-the-Wise Novgorod State University, Veliky Novgorod, 173003, Russia}
	
	\address{
		$ ^{2} $Tver State University, Tver, 170002, Russia}
	
	\ead{anatoly.zakharov@novsu.ru; victor.v.zubkov@gmail.com}

	\begin{abstract}
		
It is proved that the class of stable interatomic potentials admits an exact representation in the form of a finite or infinite superposition of Yukawa potentials. An auxiliary scalar field is introduced to describe the dynamics of a system of neutral particles (atoms) in the framework of classical field theory. In the case of atoms at rest, this field is equivalent to the interatomic potential, but in the dynamic case it describes the dynamics of a system of atoms interacting through a relativistic classical field. A relativistic Lagrangian for a system consisting of atoms and an auxiliary composite field through which the atoms interact is proposed. Equations are derived for the relativistic dynamics of a system consisting of atoms and an auxiliary field via which the atoms interact. A closed system of equations for the relativistic dynamics of a system consisting of atoms and an auxiliary field through which the atoms interact is derived. In the resulting system, an exact analytical exclusion of field variables is performed, and a closed kinetic equation is derived with respect to the probability-free microscopic atomic distribution function.

Keywords: Classical relativistic dynamics; Causality principle; Stable interatomic interactions; Irreversibility phenomenon; Retarded interactions
	\end{abstract}

\maketitle 

\section{Introduction}

Theoretical studies of the thermodynamic and kinetic properties of condensed systems are conducted mainly on the basis of statistical mechanics in the framework of the non-relativistic approximation. The usual argument in favor of the non-relativistic approximation is the relative smallness of the relativistic effects, which are of the order $  \varepsilon \sim v/c \ll 1 $, where  $ v $  is the characteristic velocity of particle motion, $  c  $ is the speed of light.

However, the most fundamental problems of thermodynamics have not been satisfactorily substantiated within the framework of statistical mechanics. Let's note some of these problems.

		\begin{enumerate}
	\item The origin of the state of thermodynamic equilibrium. Statistical mechanics is based on self-contained classical mechanics, within which thermodynamic equilibrium is impossible. Combining the mutually exclusive concepts of probability and deterministic classical mechanics makes statistical mechanics an ill-posedness theory. The hidden nonphysical concept of probability in the classical mechanics of both many-particle and few-particle systems can in no way explain the physical mechanisms of the real irreversibility of the world, both in the small and in the whole.

	\item The fundamental impossibility of microscopic substantiation within the framework of classical Newtonian mechanics. In order for an isolated system of particles to be able to move to an equilibrium state and remain in this state for an arbitrarily long time, the existence of a preferred direction of time is necessary, or, in other words, the nonequivalence of the past and the future. Today, the only version of such a theory is the special theory of relativity (STR) along with the principle of causality. The effect of STR is the impossibility of instantaneous interactions between distant particles, and the causality principle implies the existence of delayed and nonexistence of advanced interactions. Thus, within the framework of STR, there is a mechanism leading to the irreversibility of the dynamics of a system of particles.
		\end{enumerate}

Unlike classical mechanics, in which the dynamics of a system of particles completely determined by its Hamiltonian, which depends on the simultaneous coordinates and momenta of the particles, in the framework of the theory of relativity, the potential energy of interacting particles does not exist as a function of their positions.
Therefore, a correct description of the dynamics of a system of interacting particles is possible only within the framework of the field theory. As a result, the dynamics of both many-body and few-body systems should contain not only the dynamics of particles, but also the evolution of the field through which the particles interact.

A consequence of the field nature of interactions between particles is the retardation effect. 
This effect is a singular perturbation, i.e. perturbation, leading to a qualitative change in the dynamics of the system.
The study of systems with retarded interactions leads to the need to use the mathematical apparatus of functional differential equations. Due to the insufficient development of this mathematical apparatus, today it has been possible to investigate and solve only an extremely limited number of model problems, as a rule, two-body problems~\cite{Synge1,Driver1,Hsing,Driver2, Zakharov2019}. In these works, it was established that the retardation in interactions leads to the irreversibility of the dynamics of even two-body systems.

In the work~\cite{Zakharov2021}, the dynamics of free and forced oscillations of a chain of particles were investigated in a harmonic model, considering the retardation of interactions between atoms. An amazing result was established: the retardation in interactions between particles leads to the impossibility of stationary free vibrations of the crystal lattice. It is shown that in the case of a stable lattice, forced vibrations, regardless of the initial conditions, pass into a stationary regime, which is interpreted as a state of thermodynamic equilibrium in crystals.

Thus, the concept of probability is not necessary to explain both the phenomenon of irreversibility and the existence of thermodynamic equilibrium in many- and few-body systems.
To quantitatively describe these phenomena within the framework of the classical relativistic approach, it must construct a theory that includes the dynamics of particles and the evolution of the field through which the particles interact. Such a theory in the case of a system consisting of charged particles is classical electrodynamics. Within the framework of classical electrodynamics, the complete system of equations consists of the equations of particle dynamics and Maxwell's equations.
In papers~\cite{Zakharov2020-arxiv, Zakharov2020, Zubkov2020} in the complete system of equations of classical electrodynamics and equations of charges dynamics, field variables are excluded and closed systems of functional differential equations for microscopic (non-probabilistic) distribution functions of charged particles are obtained.

In works~\cite{ZakZub1, ZakZub2}, a kinetic equation was obtained for the microscopic distribution function of a system of neutral identical particles interacting through an arbitrary scalar potential considering retardation. It is proved that in the case of stable interatomic potentials~\footnote{Interatomic potentials are called stable or non-catastrophic potentials, for which the logarithm of the partition function of a system of particles in the thermodynamic limit is an extensive thermodynamic function. The stability criterion for potentials was established in the works of Dobrushin, Fisher and Ruelle~\cite{Dob-1, Fish-1, Fisher-Rue, Ruelle}.} in the asymptotics of large times $ t \to \infty $, the motion of all particles of the system stops. This solution was interpreted as an irreversible transition of the kinetic energy of particles into the energy of the field, which goes to infinity.

A detailed study of the dynamics of a system of atoms, the interaction between which at rest can be described by given interatomic potentials, requires the construction of a relativistic theory of the corresponding scalar field.

In this paper, we propose a method of classical field-theoretical description of the dynamics of a system of neutral particles (atoms).

\section{Transition from instantaneous potentials to a field picture of interactions}

To describe the dynamics of a system of neutral particles in terms of the classical field theory, we introduce an auxiliary scalar field $ \varphi\left( \mathbf{r}, t \right)$, which in the static case (i.e., in the case of particles at rest) is equivalent to the central interatomic potential $ v\left( r\right) $, and in the dynamic case allows us to describe the dynamics of a system of atoms in terms of the relativistic classical field theory.

\subsection{Static fields}

Let us assume that the interatomic potential for particles at rest admits the Fourier transform
\begin{equation}\label{v(r)}
	v\left( \mathbf{r}\right) = \int \frac{d\mathbf{k}}{\left( 2\pi\right)^{3} }\, \tilde{v}\left( \mathbf{k}\right)\, e^{-i \mathbf{k r}}.
\end{equation}
Let us denote by $ f \left(x \right) $ an ``arbitrary'' function of one variable $ x $ and apply the corresponding operator $ f \left(\Delta \right) $ ($ \Delta $ is the Laplace operator) on the function $ v \left(\mathbf{r} \right) $:
\begin{equation}\label{f(Delta)}
	f\left( \Delta\right) \left\lbrace  v\left( \mathbf{r}\right)\right\rbrace  = \int \frac{d\mathbf{k}}{\left( 2\pi\right)^{3} }\,  f\left(  -k^{2}\right)  \, \tilde{v}\left( \mathbf{k}\right)\, e^{-i \mathbf{k r}}
\end{equation}
where $ k = \left| \mathbf{k}\right|  $.
We find the explicit form of the function $ f\left( x\right)  $  from the condition (the case of a point source of the field)
\begin{equation}\label{f(delta)1}
	f\left( \Delta\right) \left\lbrace  v\left( \mathbf{r}\right)\right\rbrace  = 	\delta\left( \mathbf{r}\right).
\end{equation}
This implies
\begin{equation}\label{f(k2)}
	f\left( -k^{2}\right) = \frac{1}{\tilde{v}\left( \mathbf{k}\right)}.
\end{equation}
This relation establishes a connection between the Fourier transform of the static potential $ \tilde{v}\left( \mathbf{k}\right) $ and the differential equation~\eqref{f(delta)1}, which describes the corresponding static field. As an illustration, consider two simple examples.

\begin{enumerate}
	\item  Let 
\begin{equation}\label{Coulomb}
		\tilde{v}\left( \mathbf{k}\right) = \frac{4\pi}{k^{2}}.
\end{equation}
Then 
\begin{equation}\label{Coulomb2}
	f\left( \Delta\right) = -\frac{1}{4\pi}\, \Delta. 
\end{equation}
Hence follows the Poisson equation for the static field
\begin{equation}\label{Coulomb3}
-\frac{1}{4\pi}\, \Delta\left\lbrace \frac{1}{r} \right\rbrace = \delta \left( \mathbf{r}\right).
\end{equation}

\item Let
\begin{equation}\label{Yukawa}
		\tilde{v}\left( \mathbf{k}\right) = \frac{4\pi}{k^{2} + \mu^{2}}.
\end{equation}
Then we have
\begin{equation}\label{Yukawa2}
	f\left( \Delta\right) = -\frac{1}{4\pi}\, \left( \Delta -\mu^{2} \right). 
\end{equation}
Hence follows the Yukawa (Debye-H\"{u}ckel) equation for the static field
\begin{equation}\label{Yukawa3}
 -\frac{1}{4\pi}\, \left( \Delta -\mu^{2} \right)\left\lbrace \frac{e^{-\mu r}}{r} \right\rbrace = \delta \left( \mathbf{r}\right).
\end{equation}

\end{enumerate}

\subsubsection{Stable potentials}

Within the framework of classical statistical mechanics, only \textit{stable} interatomic potentials can be used that satisfy the criterion of Dobrushin, Fischer and Ruelle. In terms of the Fourier transform, this condition has the form~\cite{Baus}
\begin{equation}\label{Dobrushin}
\tilde{v}\left( \mathbf{k}\right) > 0	
\end{equation}
for all values of $ \mathbf{k} $.
Thus, the static equation of the scalar field $ \varphi\left( \mathbf{r}\right) $, through which interactions between atoms \textit{at rest} are conducted, will be sought in the form
	\begin{equation}\label{stat-field}
		f\left( \Delta\right) \left\lbrace  \varphi\left( \mathbf{r}\right)\right\rbrace 	= \rho\left(\mathbf{r} \right),
	\end{equation}
where  $ f\left( -k^{2}\right)  $ is the function determined by the interatomic potentials using the relation~\eqref{f(k2)}, $ \rho\left(\mathbf{r} \right)$~is the density of the field sources.

\subsubsection{The rational model of Fourier transforms of interatomic potentials}
The model concretisation consists of the choice and substantiation of explicit forms of the interatomic potentials $ v\left( \mathbf{r}\right)  $ or their Fourier transforms $ \tilde{v}\left( \mathbf{k}\right)  $, satisfying the condition~\eqref{Dobrushin}. 

First, note that in particular cases~\eqref{Coulomb} and~\eqref{Yukawa}, the Fourier transforms of potentials are:
\begin{enumerate}\label{tilde-v}
	\item proper rational functions with respect to the variable $ k $;
	\item contain only even powers of the variable $ k $;
	\item the singular points of the Fourier transforms do not lie on the real axis of the variable $ k $.
\end{enumerate}

The simple generalization of the Coulomb~\eqref{Coulomb} and Yukawa~\eqref{Yukawa} cases is a class of model interatomic potentials whose Fourier transforms are rational functions of $  k $:
\begin{equation}\label{Four-appr}
	\tilde{v}\left( \mathbf{k}\right) = \frac{Q_{m}\left( k\right) }{P_{n}\left( k\right)}, \quad (m<n),  	
\end{equation}
where
$ Q_{m}\left( k\right)$ and $ P_{n}\left( k\right) $~are polynomials of degree $ 2m $ and $ 2n $, respectively:
\begin{equation}\label{polynom}
	P_{n}\left( k\right) = \sum_{s=0}^{n}\,	C_{s}\, k^{2s},\quad Q_{m}\left( k\right) = \sum_{s=0}^{m}\,	D_{s}\, k^{2s}.
\end{equation}
$ C_{s},\, D_{s} $ are real coefficients. 
From the condition~\eqref{Dobrushin} it follows that the polynomials $ Q_{m} \left(k \right)$ and $ P_{n} \left (k \right) $ have no real roots.

If all roots of the polynomial  $ P_{n}\left( k\right)  $ are single, the Fourier transform of the interatomic potential can be represented as a~sum of $ n $ partial fraction
\begin{equation}\label{partial}
	\tilde{v}\left( \mathbf{k}\right) = \sum_{s=1}^{n} \frac{g_{s}}{k^{2}+\mu_{s}^{2}}.
\end{equation}
Consequently, the interatomic potential in the case when the Fourier transform of the interatomic potential is a rational function of $ k $ is a~sum of Yukawa-type potentials:
\begin{equation}\label{comb-Yu}
	v\left( \mathbf{r}\right) = \frac{1}{r} \sum_{s=1}^{n}\, g_{s}\,e^{-\mu_{s}r}.
\end{equation}

\subsection{General form of stable interatomic potentials}

Consider a continual generalization of the expansion~\eqref{partial}
\begin{equation}\label{cont}
\tilde{v}\left( \mathbf{k}\right) = \int\limits_{0}^{\infty} d\mu \ \frac{\Psi\left(\mu \right)}{k^{2} + \mu^{2}}. 		
\end{equation}
The problem is to solve this equation with respect to the function $ \Psi\left(\mu \right) $.

Performing the Fourier transform of this relation, we find
\begin{equation}\label{no-tilde}
{v}\left( \mathbf{r} \right)	
= \int \frac{d\mathbf{k}}{\left( 2\pi\right)^{3}}\, e^{i \mathbf{k r}} \int\limits_{0}^{\infty} d\mu \, \frac{\Psi\left(\mu \right)}{k^{2} + \mu^{2}} 
= \int\limits_{0}^{\infty} d\mu\, \Psi\left(\mu \right) \frac{e^{-\mu r}}{4\pi r},
\end{equation}
where $ r = \left| \mathbf{r}\right|. $
This implies
\begin{equation}\label{laplace}
	\int\limits_{0}^{\infty} d\mu\, \Psi\left(\mu \right) e^{-\mu r} =4\pi r\, v\left( r \right). 
\end{equation}
Thus, the function $4\pi r\, v\left( r \right) $ is the Laplace transform of the function $ \Psi\left(\mu \right) $.

From here we find $  \Psi\left(\mu \right) $:
\begin{equation}\label{f(mu)}
	 \Psi\left(\mu \right) = \frac{2}{i}\int\limits_{\gamma-i\infty}^{\gamma + i\infty} r v\left( r\right) \, e^{r\mu}\, dr, \quad \gamma > a,
\end{equation}
where $ a $~is the abscissa of absolute convergence of the Laplace transform~\eqref{laplace}:
\begin{equation}\label{a-Laplace}
	\left| r v\left( r\right) \right| \leq A \, e^{a r}.
\end{equation}
 
Note that the Yukawa potentials~~\eqref{comb-Yu} make  $ \delta $-shaped contributions to the function~$ \Psi\left(\mu \right) $
\begin{equation}\label{delta}
	v\left( \mathbf{r}\right) = \frac{1}{4\pi r} \sum_{s=1}^{n}\, g_{s}\,e^{-\mu_{s}r} \Leftrightarrow 	\Psi\left(\mu \right) = \sum_{s=1}^{n}\,g_{s}\, \delta\left(\mu - \mu_{s} \right). 
\end{equation}
However, in the general case, the support of the function  $ \Psi\left(\mu \right) $  can contain both the discrete part of the ``spectrum'' with zero Lebesgue measure, and the continuous part of the ``spectrum'' with nonzero Lebesgue measure. Certainly, in the general case, $ \tilde{v}\left( \mathbf{k}\right) $ is a transcendental function. 

\section{Composite field dynamics equation}
For the first time, the transition from the static equations of Laplace and Poisson to equations relativistic in form was carried out~(in 1867!) by Lorenz and Riemann~\cite{Lorenz,Riemann}.
The result is as follows:
\begin{equation}\label{Lapl-dAlem-}
\Delta = \sum_{j=1}^{3} \diffp[2]{}{x_{j}} \to 	\Box = \sum_{j=1}^{3} \diffp[2]{}{x_{j}} - \frac{1}{c^{2}}\, \diffp[2]{}{t}.
\end{equation}
Applying this procedure to the Yukawa potential~~\eqref{Yukawa3} leads to the Klein-Gordon-Fock equation~\cite{Klein,Fock,Gordon}:
\begin{equation}\label{Klein}
\left(\Box -\mu^{2} \right) \varphi\left(\mathbf{r},t \right) = 0.
\end{equation}
This equation is the only version of a relativistic real scalar field described by a linear partial differential equation of the second order.

In the previous section, it was shown that stable interatomic potentials~ $ v\left( \mathbf{r}\right) $ admit expansion in terms of the Yukawa-type potential system~\eqref{cont}.
If the Fourier transform of the potential $ \tilde{v}\left( \mathbf{k}\right) $ is a rational algebraic function, the integral~\eqref{cont} reduces to an expansion of the form~\eqref{partial}. Then, the interaction between atoms described by the field  $ \varphi\left(\mathbf{r}, t \right)  $, consisting of $ n $ elementary fields $ \varphi_{s}\left(\mathbf{r}, t \right)  $. Each of these elementary fields is characterized by a single parameter $ \mu_{s} $  and obeys the Klein-Gordon equation
\begin{equation}\label{Klein2}
	\hat{L}_{s} \varphi_{s}\left(\mathbf{r}, t \right) =0,
\end{equation}
where
\begin{equation}\label{Ls}
	\hat{L}_{s} = \Box - \mu_{s}^{2}.
\end{equation}
Operators $ \hat{L}_{s} $ form a set of pairwise commuting linear differential operators of the second order with constant coefficients.

We put
\begin{equation}\label{sum-phi}
	\varphi\left(\mathbf{r}, t \right) = \sum_{s=1}^{n} g_{s}\, \varphi_{s}\left(\mathbf{r}, t \right)
\end{equation}
where $ g_{s} $ are the coefficients determined by the expansion~\eqref{partial} and introduce a linear partial differential operator with constant coefficients
 \begin{equation}\label{L}
 	\hat{L} = \prod_{s=1}^{n}\hat{L}_{s}.
 \end{equation}
Since
 \begin{equation}\label{Lss}
 	\hat{L}_{s} \varphi_{s'}\left(\mathbf{r}, t \right) = \left(\mu_{{s'}}^{2}-\mu_{{s}}^{2} \right) \varphi_{s'}\left(\mathbf{r}, t \right),
 \end{equation}
then 
\begin{equation}\label{L-full}
\hat{L}	\varphi\left(\mathbf{r}, t \right) = 0.
\end{equation}
 
Thus, there are two options for describing the scalar field through which the interaction between atoms occurs. 
\begin{enumerate}
	\item Using the Klein-Gordon equation for each component $ \varphi_{s}\left(\mathbf{r}, t \right) $ and then combining their \eqref{sum-phi} into a complete field.
	\item Using full linear differential equation~\eqref{L-full} of order $ 2n $. 
\end{enumerate}
From a practical point of view, the first option is preferable, since the mathematical apparatus (both classical and quantum) of the theory of a real Klein-Gordon scalar field is quite well-developed.

\section{Retarded potentials of Yukawa fields and their superpositions}
The fundamental solution (Green's function in physical terminology) of the Klein-Gordon operator is defined by the equation
\begin{equation}\label{found-sol}
	\left(\Box -\mu_{s}^{2} \right) G_{s}\left(\mathbf{r-r'},t-t' \right) = -\delta\left(\mathbf{r-r'} \right)\, \delta\left( t-t'\right)
\end{equation}
and has the well-known form~\cite{Morse}
\begin{equation}\label{Found-Fourier}
	\begin{array}{r}
		{\displaystyle  G_{s}\left(\mathbf{r-r'},t-t' \right) = \frac{\delta\left(t-t' - \frac{\left| \mathbf{r-r}'\right| }{c}\right) }{4\pi \left| \mathbf{r-r}'\right| } }\\
		{\displaystyle  - \theta\left( t-t' -\frac{\left| \mathbf{r-r}'\right| }{c} \right)\, c\mu_{s}\, \frac{J_{1} \left( \mu_{s}\sqrt{c^{2}\left( t-t'\right) ^{2}-\left| \mathbf{r-r}'\right|^{2}}\right)}{4\pi\sqrt{c^{2}\left( t-t'\right) ^{2}-\left| \mathbf{r-r}'\right|^{2}} }, }
	\end{array}
\end{equation}
where $ \theta\left( t\right)  $~is the Heaviside step function, $ J_{1}\left( x\right)  $~is is the Bessel function. 
Hence follows the retarded potential of the Klein-Gordon field~\cite{Ivanenko} 
\begin{equation}\label{ret-Yukawa}
\begin{array}{r}
{\displaystyle  \varphi_{s}\left( \mathbf{r}, t\right) = \int d\mathbf{r}' \Biggl[\frac{\rho\left(\mathbf{r}', t-\frac{\left|\mathbf{r - r}' \right| }{c} \right) }{4\pi\left|\mathbf{r - r}' \right| }   }\\
{\displaystyle  - \mu_{s}\int\limits_{0}^{\infty}\rho\left( \mathbf{r}', t - \frac{1}{c} \sqrt{\xi^{2}+ \left|\mathbf{r - r}' \right|^{2} } \right)\, \frac{J_{1}\left(\mu_{s}\xi \right) } {4\pi\sqrt{\xi^{2}+ \left|\mathbf{r - r}' \right|^{2} } }\, d\xi \Biggr], }
\end{array}
\end{equation}
where $ \rho\left(\mathbf{r}, t \right) $  is the instantaneous microscopic density of the number of particles (atoms):
\begin{equation}\label{rho}
	\rho\left(\mathbf{r}, t \right) = \sum_{a} \delta\left(\mathbf{r} - \mathbf{r}_{a}\left(t \right)  \right), 
\end{equation}

Thus, the field $ \varphi_{s}\left( \mathbf{r}, t\right)  $  consists of two contributions.
\begin{enumerate}
	\item The first contribution determines the one-to-one relationship between the distance $ \left| \mathbf{r-r'}\right|  $ and the retardation of interactions $ \tau_{1} $ between points $ \mathbf{r,\, r'} $:
		\begin{equation}\label{tau1}
		\tau_{1} = 	\frac{\left|\mathbf{r - r}' \right| }{c}.
	\end{equation}
This contribution corresponds to a d'Alembert wave propagating from a source at the speed of light $ c $.
	\item The second contribution contains a whole spectrum of retardations in interaction between points $ \mathbf{r} $ and $ \mathbf{r}' $, depending on the parameter~$ \xi $:
	\begin{equation}\label{tau2}
		\tau_{2}\left( \xi\right) = \frac{\sqrt{\xi^{2}+ \left|\mathbf{r - r}' \right|^{2} }}{c}, \quad {0} < \xi < \infty, \quad \tau_{1} < \tau_{2}\left( \xi\right) < \infty.
	\end{equation}
This contribution corresponds to a set of Klein-Gordon waves propagating from a source with all velocities $ \tilde{c}\left(\xi \right)  $ from $ 0 $ to $ c $:
\begin{equation}\label{tilde-c}
	\tilde{c}\left(\xi \right) = c \left( 1+ \frac{\xi^{2}}{\left|\mathbf{r - r}' \right|^{2}} \right)^{-1/2},\quad  0 < \tilde{c}\left(\xi \right)  < c.
\end{equation}
Therefore, the presence of the field parameter $ \mu_{s} $ increases the retardation in interactions between particles: the function $ \tau_{2}\left( \xi\right) $ is unbounded above.
\end{enumerate}

Using relations~\eqref{partial} and~\eqref{sum-phi}, we find the connection between the dynamics of a system of atoms and the relativistically invariant auxiliary field $~\varphi\left( \mathbf{r}, t\right) $, through which the atoms interact:
\begin{equation}\label{full-field}
	\begin{array}{r}
{\displaystyle  \varphi\left( \mathbf{r}, t\right) = \int d\mathbf{r}' \sum_{s=1}^{n} g_{s} \Biggl[\frac{\rho\left(\mathbf{r}', t-\frac{\left|\mathbf{r - r}' \right| }{c} \right) }{4\pi \left|\mathbf{r - r}' \right| }   }\\
{\displaystyle  - \mu_{s}\int\limits_{0}^{\infty}\rho\left( \mathbf{r}', t - \frac{1}{c} \sqrt{\xi^{2}+ \left|\mathbf{r - r}' \right|^{2} } \right)\, \frac{J_{1}\left(\mu_{s}\xi \right) } {4\pi\, \sqrt{\xi^{2}+ \left|\mathbf{r - r}' \right|^{2} } }\,  d\xi \Biggr] . }
	\end{array}
\end{equation}
This expression has an elementary generalization to the case when the support of the function $ \Psi\left( \mu\right)  $ is not limited to the discrete spectrum:
\begin{equation}\label{full-cont}
	\begin{array}{r}
	{\displaystyle  \varphi\left( \mathbf{r}, t\right) = \int d\mathbf{r}' \int d\mu\, \Psi\left(\mu \right) \Biggl[ \frac{\rho\left(\mathbf{r}', t-\frac{\left|\mathbf{r - r}' \right| }{c} \right) }{4\pi \left|\mathbf{r - r}' \right| }   }\\
	{\displaystyle  -\mu \int\limits_{0}^{\infty}\rho\left( \mathbf{r}', t - \frac{1}{c} \sqrt{\xi^{2}+ \left|\mathbf{r - r}' \right|^{2} } \right)\, \frac{J_{1}\left(\mu\xi \right) } {4\pi\sqrt{\xi^{2}+ \left|\mathbf{r - r}' \right|^{2} } }\, d\xi \Biggr] . }
\end{array}	
\end{equation}

Thus, the instantaneous configuration of the auxiliary relativistic field $\varphi\left( \mathbf{r}, t\right)  $ at time $ t $ depends on all configurations of the atomic system described by the function $ \rho\left(\mathbf{r}, t' \right) $ under the condition  $ -\infty < t' \leq t $, i.e., on the entire prehistory of the mechanical component of the complete system, which includes both atoms and the auxiliary field created by the atoms. \footnote{Note, by the way, that even with $ \left| \mathbf{r - r} '\right| \to 0 $ the retardation $ \tau_{2} \left(\xi \right) $ reaches arbitrarily large values.}

However, just one equation~\eqref{full-cont}, which describes the effect of atoms on the field, is not enough to complete the description of the dynamics of the system as a whole: we also need equations that describe the effect of the field on atoms. 

\section{Equations of motion of particles and field}
To derive the equations of the evolution of a system consisting of particles and the field created by them, we turn to the variational formulation of dynamics. The action of the system under consideration in the case of the Yukawa field has the form
\begin{equation}\label{Action}
	\begin{array}{c}
{\displaystyle 	S=-\sum\limits_{a} m_{a}c\int \ ds_{a}-\frac{\gamma}{c}\sum_{a}\int \varphi(x_{a}) \ ds_{a} }\\
{\displaystyle +\frac{\varkappa}{2c} \int   d^{4}x\,  \left( \partial_{\nu} \varphi(x)\, \partial^{\nu}\, \varphi(x) - \mu^2\varphi^2(x) \right),}
	\end{array}
 \end{equation}
where $\gamma$ is the coupling constant between particles and the field, $\varkappa$ is a dimensional constant.

The complete system of equations for the dynamics of particles and fields has the form
\begin{equation}\label{eq-mot}
\left\lbrace 
	\begin{array}{l}
{\displaystyle   \frac{\partial}{\partial x^{\nu}} \ \frac{\partial \mathcal L}{\partial (\partial_{\nu}\varphi) } - \frac{\partial \mathcal L}{\partial \varphi}=0;}\\
{\displaystyle \frac{d}{d\tau_a} \ \frac{\partial L}{\partial \dot{x}_{a}^{\nu} } - \frac{\partial L}{\partial x_{a}^{\nu}}=0. }
	\end{array}
	\right. 
\end{equation}

For the convenience of calculations, we transform the first two terms on the right-hand side of the expression~\eqref{Action}:
\begin{equation}\label{Action1}
	\begin{array}{l}
		{\displaystyle -\sum\limits_{a} m_{a}c\int \, ds_{a}-\frac{\gamma}{c}\sum_{a}\int ds_{a}\, \varphi(x_{a}) \ }\\
		{\displaystyle = -c\sum\limits_{a} m_{a}\int \, d\tau_{a}\, \sqrt{\dot{x}^{\nu}_{a} \dot{x}_{\nu a}}-\frac{\gamma}{c}\sum_{a}\int d\tau_{a}\, \varphi(x_{a}) \, \sqrt{\dot{x}^{\nu}_{a} \dot{x}_{\nu a}} }\\
		{\displaystyle  = -c\sum\limits_{a} m_{a}\int \, d\tau_{a}\, \sqrt{\dot{x}^{\nu}_{a} \dot{x}_{\nu a}} }\\
		{\displaystyle - \frac{\gamma}{c}\sum_{a}\int d\tau_{a} \int d^4 x \ \varphi(x) \sqrt{\dot{x}^{\nu} \dot{x}_{\nu}}\, \delta ^{4}\! \left( x-x_{a} \left(\tau _{a} \right) \right)  }
	\end{array}
\end{equation}


Substituting expressions~\eqref{Action1} and~\eqref{Action} into equations~\eqref{eq-mot}, we obtain the following system of equations for the dynamics of the field and particles
\begin{equation}\label{eq-mot2}
\left\lbrace 
\begin{array}{l}
{\displaystyle  \left( \Box - \mu^2 \right) \varphi(x) = \frac{\gamma}{\varkappa} \ \sum_{a} \int d\tau_a \sqrt{\dot{x}^{\nu} \dot{x}_{\nu}}\ \delta ^{4}\! \left( x-x_{a} \left(\tau _{a} \right) \right); }\\
{\displaystyle \left(1 +\frac{\gamma}{m_ac^2}\, \varphi(x_a)\right)\frac{dp_{\mu a}}{d\tau_{a}}=\gamma\, \frac{\partial \varphi }{\partial x_{a}^{\mu}}-\gamma\, \frac{p^{\nu a}p_{\mu a}}{m_{a}^2 c^2} \, \frac{\partial \varphi}{\partial x^{\nu a}},}
\end{array}
\right. 	
\end{equation}
where $ p_{\nu a} = m_{a}u_{\nu a} $.


The first of these equations describes the dynamics of the field $ \varphi(x) $ for a given motion of particles, determined by the right-hand side of this equation:
\begin{equation}\label{rho(x)}
	 { \frac{\gamma}{\varkappa} \ \sum_{a} \int d\tau_{a} \sqrt{\dot{x}^{\nu} \dot{x}_{\nu}} \, \delta ^{4}\! \left( x-x_{a} \left(\tau _{a} \right) \right) = \rho\left( x\right),}
\end{equation}
where $ \rho\left( x\right)$ is the instantaneous microscopic density of the number of particles~\eqref{rho}.

The second of the equations~\eqref{eq-mot2} describes the dynamics of $ a $-th particle for a given evolution of the field $ \varphi(x) $ determined by the right-hand side of this equation. However, instead of describing the dynamics of a system of particles using a system of equations, each of which corresponds to a separate particle, it is more expedient to use the kinetic equation for the Klimontovich's microscopic distribution function of the system of particles as a whole
\begin{equation}\label{Klim}
	f\left(\mathbf{r},\mathbf{p},t\right) = \sum_{a} \ \delta\left( \mathbf{r}-\mathbf{r}_{a} \left(t\right)\right) \, \delta\left( \mathbf{p}-\mathbf{p}_{a} \left(t\right)\right). 	
\end{equation}


The general solution of the field evolution equation in~\eqref{eq-mot2}   is the sum of its particular solution and the general solution of the corresponding homogeneous equation. 
In the absence of external fields, the solution to this equation has the form~\eqref{ret-Yukawa}:
\begin{equation}\label{ret-Yukawa2}
	\begin{array}{r}
		{\displaystyle  \varphi\left( \mathbf{r}, t\right) = \int d\mathbf{r}' \Biggl[\frac{\rho\left(\mathbf{r}', t-\frac{\left|\mathbf{r - r}' \right| }{c} \right) }{4\pi\left|\mathbf{r - r}' \right| }   }\\
		{\displaystyle  - \mu\int\limits_{0}^{\infty}\rho\left( \mathbf{r}', t - \frac{1}{c} \sqrt{\xi^{2}+ \left|\mathbf{r - r}' \right|^{2} } \right)\, \frac{J_{1}\left(\mu\xi \right) } {4\pi \,\sqrt{\xi^{2}+ \left|\mathbf{r - r}' \right|^{2} } }\, d\xi \Biggr]. }
	\end{array}
\end{equation}

Following Klimontovich~\cite{Klimontovich-1995}, we introduce the relativistic microscopic distribution function of particles 
\begin{equation} \label{F(x,p)} 
	{\mathcal F} \left( x,p \right)=\sum _{a}\int\, d\tau _{a}\, \delta^{4}\! \left(  x-x_{a} \left(\tau_{a} \right)\right)  \, \delta^{4} \!  \left( p-p_{a} \left(\tau _{a} \right) \right).
\end{equation} 
As shown in the paper~\cite{ZakZub1}, the distribution function~\eqref{F(x,p)} satisfies the relativistic kinetic equation in the covariant form:
\begin{equation} \label{dF-dx} 
	\left( \frac{p^{\nu } }{m } \, \frac{\partial }{\partial x^{\nu } } + F^{\nu } \left(x,p\right)\, \frac{\partial  }{\partial p^{\nu } } + \frac{\partial  F^{\nu } \left(x,p\right) }{\partial p^{\nu } } \right)  {\mathcal F} \left( x,p \right) = 0,   
\end{equation} 
where $F^{\nu }$ is the four-vector of force.

Using the expression~\cite{ZakZub1} 
\begin{equation} \label{FA2(x,p)} 
	{  {\mathcal F}} \left(x,p\right)=\frac{1}{m p^{0} } \delta \left(p^{0} -\sqrt{\mathbf{p}^{2} +m^{2} c^{2} } \right)f \left(\mathbf{r,p},t\right). 
\end{equation} 
we integrate the equation~\eqref{dF-dx} over $ p^{0} $ taking into account the relation~\eqref{eq-mot2}. 
As a result, we transform the kinetic equation~\eqref{dF-dx} of a system of particles interacting through a scalar field into the following form in terms of the Klimontovich distribution function~\eqref{Klim}
\begin{equation} \label{df-dt_1} 
	\begin{array}{c} 
		{\displaystyle \left( \frac{\partial }{\partial t} 
			+\frac{c\mathbf{p}}{\sqrt{\mathbf{p}^{2} +m^{2} c^{2} } } \frac{\partial }{\partial \mathbf{r}} +\mathbf{F}\left(\mathbf{r,p},t\right)\frac{\partial }{\partial \mathbf{p}} \right) f \left(\mathbf{r,p},t\right) } \\
		{=\displaystyle \frac{3k\left(\varphi(\mathbf{r},t)\right)}{mc^2}\left(\frac{\partial \varphi(\mathbf{r},t)}{\partial t}+
			\frac{c\mathbf{p}}{\sqrt{\mathbf{p}^2+m^2c^2}}\frac{\partial \varphi(\mathbf{r},t)}{\partial \mathbf{r}}\right)f \left(\mathbf{r,p},t\right)},
	\end{array} 
\end{equation} 
where 
\begin{equation} \label{F_} 
	\begin{array}{c}
		{\displaystyle \mathbf{F}\left(\mathbf{r,p},t\right) = -\frac{mc}{\sqrt{\mathbf{p}^2+m^2c^2}} }\,k\left(\varphi(\mathbf{r},t)\right)\\ \\
		{\displaystyle \times  \left[\frac {\partial }{\partial \mathbf{r}}+\frac{\mathbf{p}}{m^2c^2}\left(\mathbf{p}\frac{\partial }{\partial \mathbf{r}}+  \frac{\sqrt{\mathbf{p}^2+m^2c^2}}{c}\frac{\partial }{\partial t}\right)\right]\varphi(\mathbf{r},t),}
	\end{array}	
\end{equation} 
\begin{equation}\label{kL}
	k\left(\varphi(\mathbf{r},t)\right) = \frac{1}{	1 +\frac{\gamma}{mc^2}\varphi(\mathbf{r},t)}.
\end{equation}
Using \eqref{rho(x)} and~\eqref{FA2(x,p)}  auxiliary field $~\varphi\left( \mathbf{r}, t\right) $ can be written in terms of Klimontovich's microscopic distribution function:
\begin{equation}\label{ret-Yukawa_a}
	\begin{array}{r}
		{\displaystyle  \varphi\left( \mathbf{r}, t\right) = \frac{c\gamma}{\varkappa}\int \frac{d^3\mathbf{p'}}{m\sqrt{\mathbf{p'}^2+m^2c^2}}\int d^3\mathbf{r}' \int d\mu\, \Psi\left(\mu \right) \Biggl[\frac{f\left(\mathbf{r}',\mathbf{p'}, t-\frac{\left|\mathbf{r - r}' \right| }{c} \right) }{4\pi\left|\mathbf{r - r}' \right| }   } \\
		{\displaystyle  - \mu\int\limits_{0}^{\infty}f\left( \mathbf{r}', \mathbf{p'}, t - \frac{1}{c} \sqrt{\xi^{2}+ \left|\mathbf{r - r}' \right|^{2} } \right)\, \frac{J_{1}\left(\mu\xi \right) } {4\pi \sqrt{\xi^{2}+ \left|\mathbf{r - r}' \right|^{2} } } d\xi \Biggr]. }
		
	\end{array}
\end{equation}

The kinetic equation~\eqref{df-dt_1} describes the irreversible evolution of a system of particles interacting through an auxiliary relativistic scalar field~$ \varphi\left (\mathbf{r},t\right)$, which in the static mode is equivalent to the interatomic potentials, and in the dynamic mode is the transmitter of interatomic interactions.

\section{Discussion}

The main principles underlying this work are as follows.
\begin{enumerate}
	\item 
	The dynamics of a system of interacting particles is described in terms of an \textbf{exact microscopic distribution function}~\eqref{Klim}, which has no probabilistic interpretation and contains the dynamics of all the particles that make up the system. Unlike statistical distribution functions, exact microscopic functions describe the dynamics of a system, but not the evolution of probabilities.
	
	An accurate and compact description of the dynamics of a particle system is provided using distribution functions of different types. Starting with the classical works of Maxwell and Boltzmann of the second half of the 19th century, statistical distribution functions with a probabilistic interpretation are intensively used and developed.
	
	In the 20th and 21st centuries, the method of statistical distribution functions was further developed in the following directions.
	
	\begin{enumerate}
	\item 
	The Bogoliubov–Born–Green–Kirkwood–Yvon (BBGKY) hierarchy in  both classical equilibrium and classical non-equilibrium statistical mechanics~\cite{Bogoliubov,Born,Kirkwood,Yvon}. This hierarchy is a system of equations describing systems of a large number of particles with direct instantaneous interactions between them. The equation for an $ s $-particle probability density function in the BBGKY hierarchy includes the $ \left( s + 1\right)  $-particle distribution function. Therefore, the BBGKY hierarchy is not closed, and one of the most difficult problems when using this method is the problem of closing the chain of equations. 
	Unfortunately, to close the BBGKY hierarchy, one has to use additional hypotheses that go beyond the self-sufficient Gibbs theory.

	\item 
	Relativistic kinetic theory and statistical mechanics. The first works on the relativistic kinetic theory appeared soon after the creation of the theory of relativity and were sporadic until the 1970s~\cite{Juttner1, Juttner2, Tetrode, Fokker, Tolman, Synge, Chernikov, Balescu,Haar}. In subsequent years, more systematic studies were conducted in this direction~\cite{Groot2, Kuzmenkov1, Trump,Schieve, Nakamura,Quevedo, Cerc1,Liboff, Hakim, Lusanna,Veres}. Note that by virtue of the well-known ``No interactions theorem''~\cite{Currie} the correct account of direct interactions between particles within the framework of the relativistic theory is possible only in the case of collisions. Therefore, the main concepts of the non-relativistic kinetic theory (the collision integral, the molecular chaos hypothesis, the concept of probability, etc.) have also been transferred into the relativistic kinetic theory. In this regard, it is appropriate to note the controversial paper of Ritz and Einstein~\cite {Ritz}, in which Ritz hypothesized that the phenomenon of irreversibility due to the retardation of interactions between particles. Einstein in the same paper expressed the conviction that irreversibility has a probabilistic nature. 

	\item 
	The method of equilibrium and nonequilibrium Green's functions in quantum statistical mechanics~\cite{Matsubara, Martin, Bonch, Abrikosov, Kadanoff, Buh1, Buh2}.
	Quantum statistical mechanics is based on the formal analogy noted by Matsubara~\cite{Matsubara} between the classical statistical operator $ e^{-\beta H\left( q, p\right) } $ ($ \beta~=~1/\kappa T $, $ T $ is temperature, $ H \left(q, p \right) $ is the Hamiltonian of the system) and the evolution operator in quantum mechanics $ e^{-it \hat{H}} $.
	Thus, it seems that the problem of the origin of thermodynamic equilibrium remains outside the scope of quantum statistical mechanics.
	\end{enumerate}

	\item 	
	The description of particle dynamics is based on \textbf{relativistic equations of motion} and \textbf{the principle of causality}. In contrast to classical non-relativistic mechanics, in the theory of relativity there is an asymmetry between the past and the future, which gives hope for establishing a connection between the laws of thermodynamics and the laws of relativity theory.
	
	\item 	
	Interaction between atoms within the framework of the theory of relativity is possible only based on \textbf{field concepts}. Due to the neutrality of atoms, an auxiliary scalar field is used to describe inter-atomic interactions. In the case of atoms at rest, this field is equivalent to inter-atomic potentials. Thus, within the framework of the relativistic theory, the description of the dynamics of a system of interacting atoms includes both the equations of motion of particles and the equations of evolution of a scalar field that transmits interactions between particles. 
	
	\item  	
	It is proved that in the case of stable interatomic interactions, the auxiliary scalar field is a \textbf{sum of Yukawa fields}, whose parameters are uniquely expressed in terms of the interatomic potentials of atoms at rest. 
	
	As a result, the system of interacting particles consists of two subsystems, one of which is the particles, and the second is the field. The Hamiltonians of these subsystems do not exist, since the subsystems of Hamiltonian systems, generally speaking, are not Hamiltonian~\cite{Uchaikin}.
	
	\item 
	A closed \textbf{exact probability-free kinetic equation}~\eqref{df-dt_1} is derived within the framework of the theory of relativity for a system of interacting particles. 
	Note that a variant of the relativistic kinetic equation for a system of particles interacting through a field $ \varphi(\mathbf{r},t) $ with zero mass is proposed in our work~\cite{ZakZub1}. The previous version of the relativistic kinetic equation follows from the equation~\eqref{df-dt_1} if the interaction energy of the particle with the field  $ \gamma\varphi(\mathbf{r},t) $ is much less than the particle's energy~$ mc^2 $ ($ {\gamma}\varphi(\mathbf{r},t)\ll {mc^2} $).

	\item  
	The relativistic kinetic equation~\eqref{df-dt_1} is equally applicable to both many-body and few-bode systems. At the same time, in both cases, the system's dynamics have characteristic signs of its thermodynamic behavior, including both the property of irreversibility~\cite{Zakharov2019, Zakharov2020, ZakZub1} and the implementation of the microscopic equilibration mechanism~\cite{ZakZub2, Zakharov2021}.
	
Therefore, the relativistic field approach to describing the dynamics of systems can be used as a probability-free basis for constructing microscopic thermodynamics of both macroscopic and ``small'' systems, including nanosystems.

\end{enumerate}

\section{Conclusion}

We would like to express our gratitude to Ya. I. Granovsky and V. V. Uchaikin for intensive and fruitful discussions of the work.

\end{document}